%% file: 0.main.tex
\documentclass[acmsmall]{acmart} 

\AtBeginDocument{%
  \providecommand\BibTeX{{%
    \normalfont B\kern-0.5em{\scshape i\kern-0.25em b}\kern-0.8em\TeX}}}

\setcopyright{acmcopyright}
\copyrightyear{2018}
\acmYear{2018}
\acmDOI{}

\usepackage{booktabs} 
\usepackage{url}

\usepackage{caption}
\usepackage[all]{nowidow}
\usepackage{wrapfig}
\usepackage{array}
\usepackage{arydshln}
\usepackage{tabularx}
\usepackage{multirow}
\usepackage{arydshln}
\newcolumntype{L}[1]{>{\raggedright\let\newline\\\arraybackslash\hspace{0pt}}m{#1}}
\newcolumntype{C}[1]{>{\centering\let\newline\\\arraybackslash\hspace{0pt}}m{#1}}
\newcolumntype{R}[1]{>{\raggedleft\let\newline\\\arraybackslash\hspace{0pt}}m{#1}}

\usepackage{wrapfig,lipsum,booktabs} 

\usepackage{pgfplots}
\pgfplotsset{compat=1.18}

\def\authnotes{1}
\newcounter{notectr}[section]
\newcommand{\thenote}{\thesubsection.\arabic{notectr}\refstepcounter{notectr}}

\newcommand{\note}[2]{$\ll$#1~\thenote: #2$\gg$}
\newcommand{\cnote}[1]{\ifnum\authnotes=1 \textcolor{blue}{\note{Comment:}{#1}}\fi}

\acmYear{2026}
\copyrightyear{2026}

\begin{document}


\title[short title]{ "OpenBloom": A Question-Based LLM Tool to Support Stigma Reduction in Reproductive Well-Being }

\author{Ashley Hua}
\authornotemark[1]
\affiliation{%
  \institution{University of Illinois Urbana Champaign}
  \city{Champaign}
  \state{Illinois}
  \country{United States}}
\email{ahhua2@illinois.edu}

\author{Adya Daruka}
\authornotemark[1]
\affiliation{%
  \institution{University of Illinois Urbana Champaign}
  \city{Champaign}
  \state{Illinois}
  \country{United States}}
\email{adaruka2@illinois.edu}

\author{Yang Hong}
\affiliation{%
  \institution{ University of Illinois Urbana Champaign}
  \city{Champaign}
  \state{Illinois}
  \country{United States}}
\email{yangh9@illinois.edu}

\author{Sharifa Sultana}
\affiliation{%
  \institution{ University of Illinois Urbana Champaign}
  \city{Champaign}
  \state{Illinois}
  \country{United States}}
\email{sharifas@illinois.edu}

\renewcommand{\shortauthors}{Hua et al.}

\begin{abstract}
Reproductive well-being education remains widely stigmatized across diverse cultural contexts, constraining how individuals access and interpret reproductive health knowledge. We designed and evaluated OpenBloom, a stigma-sensitive, AI-mediated system that uses LLMs to transform reproductive health articles into reflective, question-based learning prompts. We employed OpenBloom as a design probe, aiming to explore the emerging challenges of reproductive well-being stigma through LLMs. Through surveys, semi-structured interviews, and focus group discussions, we examine how sociocultural stigma shapes participants’ engagements with AI-generated questions and the opportunities of inquiry-based reproductive health education. Our findings identify key design considerations for stigma-sensitive LLM, including empathetic framing, inclusive language, values-based reflection, and explicit representation of marginalized identities. However, while current LLM outputs largely meet expectations for cultural sensitivity and non-offensiveness, they default to superficial rephrasing and factual recall rather than critical reflection. This guides well-being HCI design in sensitive health domains toward culturally grounded, participatory workflows.
\end{abstract}







\settopmatter{printfolios=true}

\maketitle

\input{1.intro}
\input{2.lit}
\input{3.methods}
\input{4.find}

\input{5.discussion}
\input{6.conclusion}



\bibliographystyle{ACM-Reference-Format}
\bibliography{citation.bib}

\appendix
\input{7.appendix}

\end{document}

%% file: 1.intro.tex
\section{Introduction}
Reproductive wellbeing, including menstruation, contraception, fertility, and sexual health, remains stigmatized across many communities~\cite{chowdhury2025rewa}. Cultural taboos, moral judgment, and misinformation often constrain how reproductive health knowledge is shared and interpreted, particularly among adolescents and young adults~\cite{sultana2025socheton}. These stigmas manifest in everyday practices, such as silence around menstruation in schools, shame surrounding contraceptive use, and discomfort discussing sexual health openly. Public health research has shown that stigma is not merely attitudinal but structurally constrains access to care; for example, unmarried women may avoid reproductive health services due to fear of social judgment and discrimination~\cite{mohammadi2016stigma}. Despite sustained public health efforts, challenges persist in how reproductive health information is communicated and experienced across diverse cultural contexts.

Recent advances in AI and large language models (LLMs) introduce new forms of interaction for health education. LLMs can generate explanations, prompts, and questions that adapt to user-provided content, positioning them as potentially influential mediators of sensitive knowledge. Although the broader public health crisis surrounding health misinformation is identified as a significant threat to individual and societal well-being~\cite{surgeongeneral2021misinformation}, how LLMs frame reproductive well-being, and how their language, assumptions, and interactional styles intersect with stigma and cultural norms, remains underexplored. In reproductive health contexts, AI-generated content may reinforce existing stigma, normalize silence, or misalign with users’ expectations of sensitivity and care ~\cite{deva2025integrating}.

Prior work in HCI on reproductive wellbeing has focused on areas such as menstrual tracking, adolescent sexual health education, reproductive justice, and data privacy~\cite{chowdhury2025rewa}, while human-AI(HAI) research has examined issues of accuracy, bias, explainability, and conversational design in AI systems. Yet, limited work has examined how AI-generated prompts themselves become sites where stigma, cultural norms, and moral boundaries are interacted and negotiated, particularly in culturally sensitive health domains.

In this paper, we approach the LLM tool not as a solution to reproductive health stigma, but as a design probe for examining how stigma-sensitive interactions break down in practice. We explore the following research questions:

\begin{enumerate}
    \item What stigmas surface when participants engage with LLM-generated questions about reproductive wellbeing in diverse cultural contexts? 
    \item How do participants interpret and engage with LLM-generated questions that challenge these stigmas and foster open dialogue?
    \item How might we design LLM tools that challenge reproductive stigma while remaining culturally appropriate and educationally effective?
\end{enumerate}

To investigate these questions, we developed OpenBloom, an Android application that serves as a data collection platform and design probe. OpenBloom uses OpenAI’s LLMs to transform user-submitted reproductive health articles into AI-generated educational questions. Participants interact with these questions and reflect on their perceived relevance, clarity, creativity, and cultural sensitivity. Through surveys, interviews, and focus group discussions, we analyze how users respond to these AI-generated interactions and where they experience discomfort, misalignment, or uncertainty.

Rather than evaluating whether OpenBloom reduces stigma, our goal is to characterize the interactional limits of current LLM-generated educational content. By foregrounding moments of tension and breakdown, we surface design-relevant insights about why non-offensive language alone is insufficient for stigma-sensitive AI design. These findings highlight the need for culturally grounded, participatory design approaches that attend to how AI systems ask in sensitive health contexts.

This work contributes to HCI and HAI design scholarship by providing empirically grounded design knowledge on how users experience AI-generated questions about reproductive well-being. By using LLMs as design probes, we articulate interactional challenges and design tensions that inform future work on stigma-sensitive, culturally aware AI systems in health education.

%% file: 2.lit.tex
\section{Related Work}

Our work builds on research in AI for sexual and reproductive health education, culturally sensitive technology design, automatic question generation, and HCI scholarship on reproductive health stigma. We synthesize this literature to identify gaps motivating our investigation of how LLM-generated educational content navigates tensions between stigma reduction and cultural appropriateness.

\subsection{AI in Sexual and Reproductive Health}

AI shows promise for expanding SRH access, particularly for marginalized populations facing barriers to traditional healthcare. Systematic reviews show chatbots can provide judgment-free information addressing geographic isolation and stigma ~\cite{mills2023chatbots}, with demonstrated effectiveness across knowledge, attitudes, and self-efficacy ~\cite{fetrati2024chatbots}. However, persistent concerns about accuracy, user trust, and implementation challenges ~\cite{barreda2025transforming} suggest effectiveness depends on human support and cultural tailoring rather than standalone solutions ~\cite{aggarwal2023chatbots}.

Successful deployments demonstrate the importance of culturally grounded approaches. Nthabi in Lesotho improved family planning knowledge through stakeholder engagement ~\cite{nkabane2024nthabi}, SnehAI in India reached youth through Hindi support and anonymity ~\cite{wang2021snehAI}, and adolescents in Bangladesh valued chatbots' judgment-free nature ~\cite{rahman2021adolescentbot}. Mai for menstrual health shows promise normalizing stigmatized processes ~\cite{mughal2024mai}, though limitations in recognizing when medical consultation is necessary remain.

Critical challenges emerge in AI-generated content. Rapidly changing reproductive rights laws create accuracy issues ~\cite{bull2024feasibility}, while systematic biases produce concerning distortions: GPT's safety guardrails inadvertently generate medically inappropriate content by avoiding evidence-based pharmacological interventions ~\cite{hanai2024generative}. This "safety washing" reveals how AI systems designed to prevent harm may produce less safe, less accurate content. Addressing these challenges requires transparency, community involvement, and recognition that AI cannot substitute for systemic healthcare improvements ~\cite{antoniak2024nlp}.

Existing research focuses on information delivery rather than AI's capacity to challenge misconceptions without triggering defensive reactions—a critical gap our study addresses.

\subsection{Cultural Sensitivity in Health AI}

Culturally responsive design is essential yet difficult to implement. Frameworks articulate that AI systems must account for local norms, power dynamics, health beliefs, and structural inequities ~\cite{who2024ai}, requiring co-design with marginalized communities ~\cite{nadarzynski2024equity}. However, LLMs' linguistic fluency can mask cultural incompetence, generating natural-sounding responses reflecting inappropriate dominant cultural assumptions ~\cite{jo2023understanding}.

Empirical studies expose implementation gaps. An LLM chatbot for reproductive health in urban India struggled with culturally nuanced scenarios despite extensive prompt engineering, providing individualistic contraception advice that ignored family-centered decision-making ~\cite{deva2025integrating}. This reflects deeper issues in LLMs' cultural knowledge for non-Western contexts underrepresented in training data. Nthabi's success demonstrates what sustained community partnership can achieve ~\cite{nkabane2024nthabi}, though such approaches remain difficult to scale. The intersection of culture, religion, and reproductive health adds complexity, as beliefs are deeply embedded in religious teachings and community norms ~\cite{arousell2016culture}, requiring respectful engagement with diverse value systems ~\cite{mustafa2021religion}.

Most systems engage culture superficially through translation rather than deeper frameworks, with standardized architectures conflicting with contextual adaptation ~\cite{davies2024culturalsensitivity}. LLMs exhibit demographic disparities, producing Western-centric recommendations while generating inaccuracies and stereotypes for marginalized groups ~\cite{omar2025evaluating}, including consistently less accurate, more stereotyped content for non-white patients ~\cite{hanna2025assessing}, systematic racial bias in psychiatric recommendations ~\cite{bouguettaya2025racial}, and demographic influence on oncology recommendations despite identical clinical characteristics ~\cite{agrawal2024fairness}. In emergency care, LLMs generate biased clinical recommendations based solely on patients' sociodemographic characteristics, with Black, unhoused, or LGBTQIA+ patients receiving disproportionate mental health evaluations and invasive interventions while high-income patients receive advanced imaging ~\cite{omar2025chatbias}.

No existing research examines how marginalized communities perceive the cultural appropriateness of AI-generated educational questions about reproductive wellbeing—precisely what our study investigates.

\subsection{AI-Generated Questions for Learning}

AQG systems represent a promising but underexamined approach for health education about stigmatized topics. Recent advances show LLMs with structured prompts can generate questions meeting educational standards across cognitive levels, sometimes rivaling human-created questions ~\cite{wang2024exploring}. Yet technical capability raises a crucial question: can AI capture the pedagogical strategies that make question-based learning effective for sensitive topics?

Question-based pedagogies have demonstrated effectiveness for challenging preexisting beliefs. Socratic questioning promotes critical thinking and belief re-examination, though it requires careful facilitation to avoid defensive reactions ~\cite{burns2016socratic}. Problem-based learning consistently improves critical thinking and long-term retention by engaging learners as active problem-solvers ~\cite{polyzois2010problem}. A randomized controlled trial in Tanzania found problem-based pedagogy using questions significantly improved adolescents' reproductive health decision-making and comfort discussing sensitive topics compared to didactic teaching ~\cite{millanzi2022effect}, with teachers observing that framing content as questions allowed engagement without shame.

However, current AI applications lack this sophistication. ChatGPT for reproductive health literacy produced grammatically correct but pedagogically superficial questions defaulting to factual recall rather than engaging myths or social dimensions ~\cite{burns2024generativeai}, with users finding content "sanitized" and disconnected from community concerns. Safety mechanisms produce "safety washing" that generates less accurate content by avoiding evidence-based interventions ~\cite{hanai2024generative}.

No prior work examines whether AI-generated questions can employ strategies that reduce defensiveness about stigmatized topics or adapt to cultural contexts beyond translation—gaps our study addresses.

\subsection{Reproductive Health Stigma and HCI}

HCI research increasingly recognizes reproductive health as critical yet underexplored. The ReWA framework identifies six reproductive wellbeing activities (information seeking, tracking, decision-making, care-seeking, social support, and advocacy), showing existing technologies focus on these separately and ignore experiences beyond cisgender women planning pregnancy ~\cite{chowdhury2025rewa}.

Empirical research documents multidimensional reproductive health stigma. Ethnographic work in Bangladesh reveals stigma manifesting as internalized shame, interpersonal judgment, and structural discrimination ~\cite{chowdhury2024ancient}. Evidence from 77 studies confirms effective stigma reduction requires multi-level interventions, though technology-based approaches remain limited ~\cite{bohren2022strategies}. Stigma increases misinformation susceptibility by creating information avoidance, while information overload reinforces stigmatizing narratives ~\cite{dong2023counteracting}, suggesting educational interventions must address stigma reduction alongside information provision.

Successful interventions demonstrate culturally appropriate design's importance. Menstrupedia's visual narrative normalized menstrual health in India by engaging stakeholders and reframing menstruation as normal ~\cite{tuli2018menstrupedia}, though scaling remains challenging.

Critical feminist HCI scholarship provides essential frameworks while challenging technological solutionism. Positioning intimate care as central responds to HCI's historical neglect of reproductive health ~\cite{almeida2016hci}. Gender essentialism in health technologies marginalizes transgender and gender-diverse people while reinforcing normative assumptions ~\cite{keyes2020reimagining}. Effective design requires recognizing users as whole people navigating complex social, emotional, and political dimensions ~\cite{almeida2020woman}, with holistic, intersectional approaches ~\cite{kumar2020taking}.

Specific design implications include period-positive principles ~\cite{woytuk2019period}, critical methods revealing normative assumptions ~\cite{reime2023walking}, and documentation of privacy concerns in restrictive contexts ~\cite{mehrnezhad2023user}.

Bardzell's Feminist HCI framework proposes six qualities characterizing feminist interaction design: pluralism, participation, advocacy, ecology, embodiment, and self-disclosure ~\cite{bardzell2010feminist}. These qualities are particularly relevant for reproductive health technologies where agency, identity, equity, and social justice intersect with stigmatized embodied experiences. Our investigation directly engages with whether LLMs can support pluralist approaches, embody advocacy by challenging stigma, and achieve self-disclosure by rendering visible their assumptions about users.

While HCI scholarship offers robust frameworks for stigma-aware design, limited research examines tensions that emerge when AI-generated educational content is deployed in this complex terrain. Our study extends HCI's attention to reproductive wellbeing into generative AI by investigating user perceptions across cultural sensitivity, pedagogical effectiveness, and stigma reduction.

%% file: 3.methods.tex
\section{Methods}

\subsection{Application Design and Functionality}

Building on prior work that integrates reproductive care with LLM applications \cite{sultana2025socheton, antoniak2024nlp, jo2023understanding}, we developed a website application called OpenBloom using Flutter and Dart, serving as the data collection platform. We chose Flutter for its cross-platform compatibility and Dart for its efficient handling of asynchronous operations required for OpenAI API calls, enabling rapid prototyping while maintaining native-like performance on Android devices.

The application workflow begins with user input: participants submit an article related to a reproductive well-being issue of their choice, allowing exploration of diverse topics and sources. The app embeds the full article within the interface, enabling users to read the material directly. It then uses OpenAI's LLM to generate a concise, accessible summary, helping users better grasp the core message.

Using the article as context, the LLM generates a series of multiple-choice questions designed to test understanding and promote deeper engagement with the topic. These questions emphasize key points and encourage reflection on issues related to stigma, health literacy, and social perceptions. Users answer questions in-app and receive immediate feedback, creating an interactive learning experience.

This design allows us to collect real-time data on how users engage with AI-generated content across various reproductive health topics, providing insights into the appropriateness, sensitivity, and educational value of LLM outputs in this domain.

\begin{figure}[t!]
    \centering
    \includegraphics[width=1\linewidth]{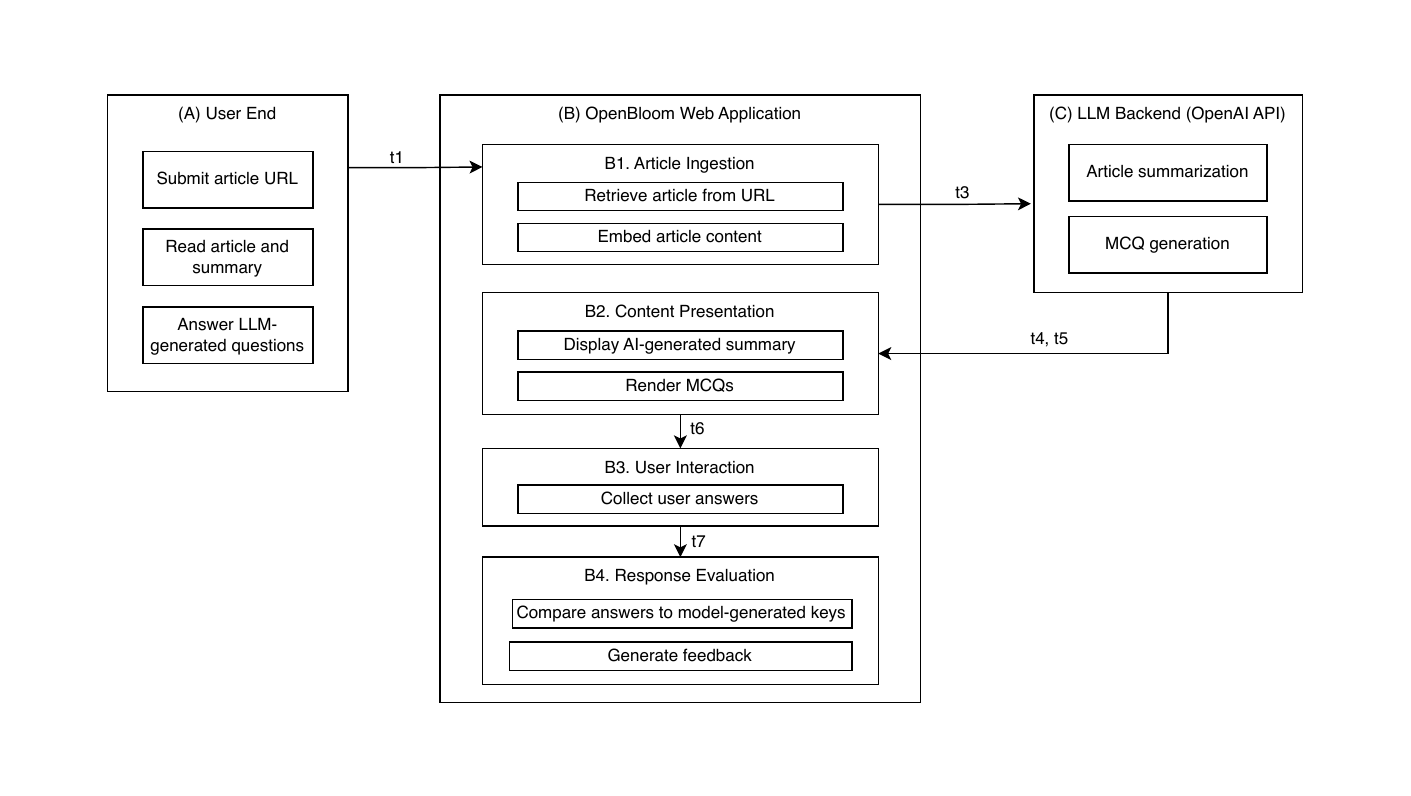}
    \caption{Work-flow diagram of OpenBloom, from article submission to AI-generated summarization, question generation, and response grading.}
    \label{fig:workflow}
\end{figure}

\begin{figure}[t!]
    \includegraphics[width=.7\textwidth]{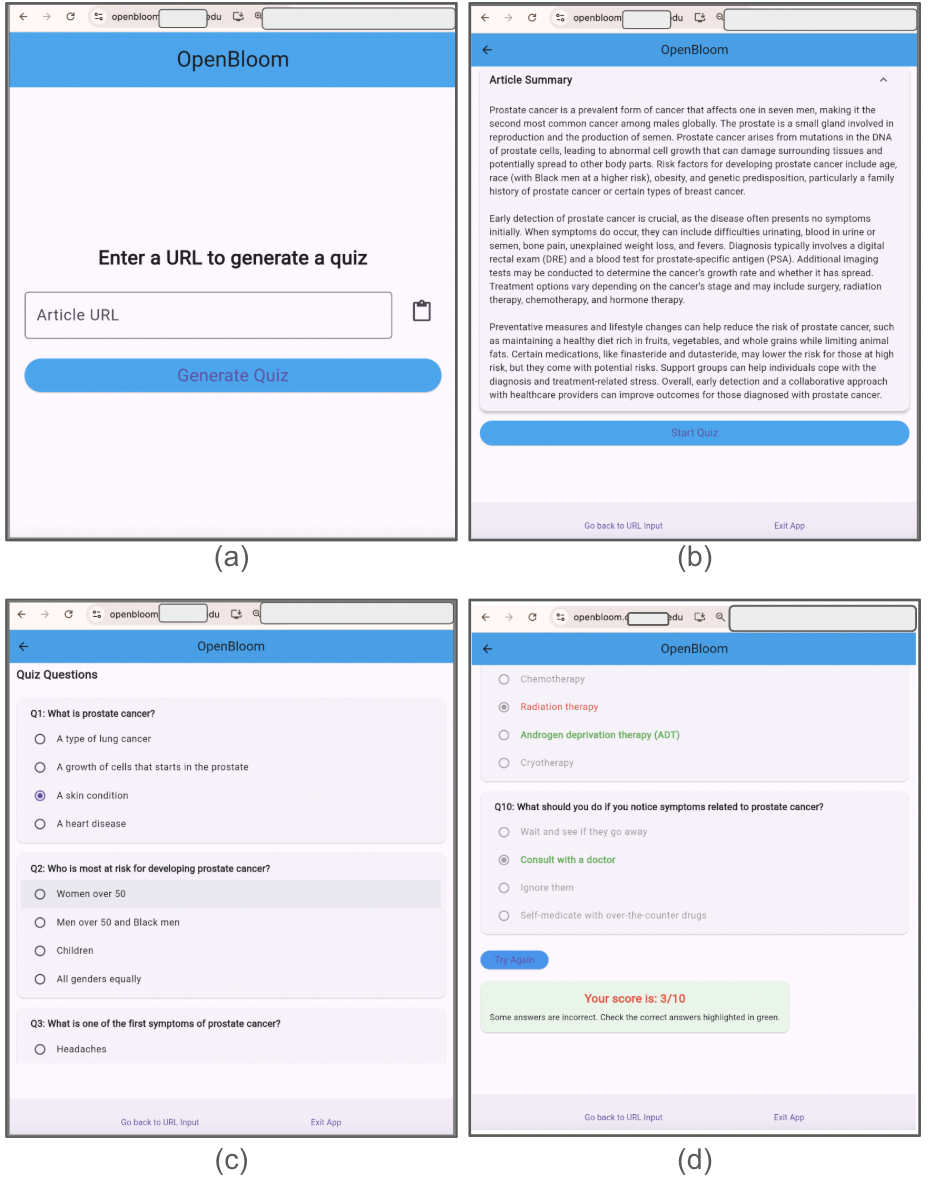}
    \caption{(a) Landing Page for OpenBloom, (b) Article summary generated by OpenBloom, (c) Excerpt of questions generated by OpenBloom, and (d) Scores shown upon taking the quiz.}
\end{figure}

As shown in Figure~\ref{fig:workflow}, the system workflow proceeds as follows:
\textbf{t1:} The user submits a reproductive health related article URL to the web application;
\textbf{t2:} the web application retrieves and embeds the article content;
\textbf{t3:} the article text is sent to the LLM backend;
\textbf{t4:} the LLM generates and returns a summary;
\textbf{t5:} the LLM generates multiple-choice questions (MCQs);
\textbf{t6:} the user answers the AI-generated questions; and
\textbf{t7:} the web application evaluates responses and provides feedback.

\subsection{Study Procedures}
Data were collected from 34 participants across two phases: (1) surveys and interviews conducted immediately after participants interacted with OpenBloom, and (2) follow-up FGDs involving eight participants, divided into two groups based on their availability. With participants’ consent, all interviews were audio-recorded and transcribed verbatim. The FGDs were not audio-recorded; the facilitating researchers documented the discussions through detailed field notes. All procedures and protocols received approval from the university’s Institutional Review Board.

We recruited participants to engage with OpenBloom through a community-wide email invitation sent to university students. In total, 34 participants completed the post-interaction interviews and surveys, and eight of them also participated in follow-up focus group discussions (FGDs). Our participants consisted primarily of college students aged 18–29, the majority of whom identified as women. Table~\ref{tab:demographics_combined} summarizes participants’ demographic characteristics.

\begin{table}[h]
\centering
\caption{Demographics of the participants in interviews and focus group discussions (FGDs) (N=34)}
\label{tab:demographics_combined}
\begin{tabular}{lcc}
\toprule
\textbf{Characteristic} & \textbf{Category} & \textbf{N (\%)} \\
\midrule
\multirow{2}{*}{Gender} & Female & 27 (79\%) \\
                        & Male & 7 (21\%) \\
\midrule
\multirow{3}{*}{Type of Participation} & Interview only (P1-P30) & 26 (76\%) \\
                                       & FGD only (initial) & 4 (12\%) \\
                                       & Interview + FGD & 4 (12\%) \\
\midrule
\multirow{4}{*}{Age (years)} & 18-19 & 20 (59\%) \\
                             & 20-22 & 10 (29\%) \\
                             & 25-26 & 3 (9\%) \\
                             & 29 & 1 (3\%) \\
                             & Mean (SD) & 20.1 (2.6) \\
\midrule
\multirow{6}{*}{Race} & Indian & 16 (47\%) \\
                      & Chinese & 8 (24\%) \\
                      & American & 7 (21\%) \\
                      & White & 1 (3\%) \\
                      & Norwegian & 1 (3\%) \\
                      & Sri Lankan & 1 (3\%) \\
\midrule
\multirow{6}{*}{Occupation} & Student & 26 (76\%) \\
                            & Software Engineer & 3 (9\%) \\
                            & Consultant & 2 (6\%) \\
                            & Product Manager & 1 (3\%) \\
                            & Sales Associate & 1 (3\%) \\
                            & Manager & 1 (3\%) \\                     
\midrule
Status & Undergraduate/Graduate students & 34 (100\%) \\
\bottomrule
\end{tabular}
\end{table}

\subsubsection{Survey}
Participants were asked to interact with the application using four different articles of their choice. For each article, they completed a quantitative survey measuring their perceptions of the AI-generated questions across four dimensions: creativity, persuasiveness, relevance, and cultural sensitivity. All questions used a 5-point Likert scale (Strongly Agree to Strongly Disagree). Survey questions are provided in the Appendix.

\subsubsection{Individual Interviews}
Immediately following their interaction with the application and survey completion, participants engaged in semi-structured individual interviews. These interviews, lasting 10-30 minutes depending on participant detail, explored their reactions to specific questions, comfort level discussing various reproductive health topics, and perceptions of the AI's cultural sensitivity and appropriateness.

Importantly, surveys and interviews were conducted simultaneously for each participant rather than sequentially across the participant pool. This approach captured immediate reactions while participants' experiences with the application remained fresh.

\subsubsection{Focus Group Discussions}
Based on preliminary analysis of survey responses and interview transcripts, we designed focus group discussion protocols to explore emergent themes in greater depth. These sessions brought together eight participants from diverse backgrounds to discuss their reactions to specific AI-generated content, explore cultural variations in stigma perceptions, and identify patterns in what makes AI-generated health communication effective or problematic.

Focus groups allowed participants to build on each other's observations, revealing nuanced insights about cultural sensitivity, appropriate language choices, and the potential role of AI in stigma reduction that might not emerge in individual interviews alone.

\subsection{Data Analysis}

We collected survey responses from 34 participants across 136 article interactions (4 articles per participant), interview transcripts totaling approximately 53 pages of field notes, and focus group recordings and transcripts.

For quantitative data, we conducted descriptive statistical analysis to identify patterns in participant ratings across the four dimensions (creativity, persuasiveness, relevance, cultural sensitivity) and examine variations based on article topics and participant demographics.

For qualitative data analysis, we employed reflexive thematic analysis following Braun and Clarke's approach~\cite{braun2006using}. Two researchers independently coded interview and focus group transcripts using open coding to identify initial patterns. Through iterative discussion and refinement, we developed a coding framework that captured key themes related to: (1) perceptions of AI-generated content quality, (2) cultural appropriateness and sensitivity, (3) educational effectiveness, and (4) concerns about AI in reproductive health education.

By combining quantitative survey data with qualitative insights from interviews and focus groups, we aimed to develop comprehensive, evidence-based guidelines for how LLMs should be designed and deployed to discuss reproductive well-being topics sensitively across different cultural contexts. This multi-method approach allowed us to capture both measurable user perceptions and the rich, contextual understanding necessary for developing culturally-aware health communication systems.

%% file: 4.find.tex
\section{Findings}

Our findings draw on focus group discussions (N=8 participants across 2 groups) and individual interviews (N=30), exploring reproductive health stigma, pedagogical approaches for AI-generated content, and design principles with user evaluations. Focus group participants are identified by their initials: FG1-A, FG1-M, FG1-C, and FG1-E (Focus Group 1) and FG2-J, FG2-R, FG2-S, and FG2-L (Focus Group 2). Interview participants are identified by participant ID (P1-P30).

\subsection{Reproductive Health Stigma: Manifestations, Myths, and Marginalization}

Five primary stigma categories emerged from our analysis: menstruation, contraception, abortion and STI (sexually transmitted infection) care, fertility and infertility, and media-driven body image concerns. These manifestations encompass both individual experiences of shame and systemic healthcare barriers.

\subsubsection{Different Types of Stigma}

Menstruation stigma begins early in development through gendered health education practices that exclude boys from learning about menstrual health, perpetuating ignorance that later manifests as stigma (FG1-A). The consequences extend beyond psychological discomfort - silence surrounding menstruation delays diagnosis and treatment of conditions such as PCOS (Polycystic Ovary Syndrome), while menstrual pain becomes routinely dismissed in medical settings.

Contraception stigma operates through distinct moral frameworks rather than being conceptualized as a health tool. Three focus group participants described how contraception becomes entangled with judgments about sexual behavior and faces cultural and religious barriers (FG1-C, FG1-A, FG2-R), transforming what should represent reproductive autonomy into a source of shame.

Abortion and STI testing carry the most severe stigma. Four focus group participants noted that strong opinions about abortion access often exist without understanding the complex medical, economic, and personal factors involved (FG2-J, FG2-R, FG2-S, FG2-L). STI testing faces particular challenges, as healthcare language reinforces moral judgments - terms like "dirty" and "clean" create shame precisely when patients need support (FG2-R). Even healthcare providers often treat STI testing as taboo, causing patients to feel judged rather than supported.

Fertility stigma reveals stark gender asymmetries despite infertility affecting both partners equally. Two focus group participants emphasized how men remain systematically excluded from conversations about contraception, fertility, and pregnancy support, reinforcing the framing of reproductive health as exclusively women's responsibility (FG2-S, FG2-R). Media compounds these issues by creating unrealistic body expectations during pregnancy.

\subsubsection{Why Myths Persist Despite Available Information}

Reproductive health myths persist not from ignorance but from cultural traditions, limited historical medical knowledge, and inadequate sex education. Understanding these underlying factors proves critical for intervention design.

Five focus group participants identified numerous persistent myths across multiple domains: pregnancy timing and contraception effectiveness, hormonal side effects, virginity and the hymen, diagnosis of reproductive conditions, and which populations require reproductive healthcare (FG1-A, FG1-M, FG1-C, FG1-E, FG2-R). Each myth produces specific harms, from delaying medical care to reinforcing patriarchal control over women's bodies.

The relationship between cultural meaning and scientific accuracy proves remarkably complex. One participant shared a particularly illustrative example of her mother's instruction to wash her hair during menstruation because periods make one "dirty." Despite recognizing this belief as scientifically unfounded, she explained:

\begin{quote}
Even knowing it's untrue, I still do it as a cultural practice. (FG1-A)
\end{quote}

This reveals how few individuals interpret such beliefs literally in contemporary contexts, yet they persist socially, carrying meaning beyond their factual content.

Virginity myths produce particularly serious consequences. Three focus group participants emphasized how these myths create anxiety, reinforce patriarchal ideas about purity, and lead to harmful practices including virginity testing (FG2-L, FG2-J, FG2-R). Medical dismissal of women's pain perpetuates myths about endometriosis and PCOS - conditions that are relatively common but frequently underdiagnosed because menstrual pain becomes normalized and women's pain often lacks credibility in medical settings.

Critically, all eight focus group participants emphasized approaching myths with compassion rather than condemnation. The consensus framed myths as products of limited access to comprehensive sex education and cultural traditions rather than malicious intent, repositioning myth correction from judgment to education.

\subsubsection{Systemic Exclusion of Marginalized Communities}

Beyond individual stigma experiences, healthcare and education systems systematically exclude certain populations through both overt barriers and subtle design choices that accumulate to create unwelcoming environments.

Medical research has historically centered male subjects, creating knowledge gaps in women's health. Yet even within women's health research, cisgender experiences dominate while transgender and nonbinary individuals who menstruate, become pregnant, or need gynecological care face systems that fail to recognize their needs (FG2-L). Sex education curricula typically assume heterosexual and cisgender students, focusing predominantly on reproduction rather than diverse experiences.

Healthcare infrastructure signals belonging through seemingly minor design choices. One focus group participant described the cumulative impact of exclusionary forms:

\begin{quote}
[Healthcare forms] asked for 'husband's name' or only offered 'male' and 'female' as gender options. These small design choices accumulate to signal 'this place isn't for you.' (FG2-L)
\end{quote}

These cumulative exclusions create barriers for LGBTQ+ individuals seeking reproductive healthcare.

Disability remains systematically erased from reproductive health discourse, with disabled individuals overlooked in reproductive education as though they are not sexually active. This erasure extends to digital spaces where applications and websites often lack adequate alternative text, captions, and appropriate color contrast.

Intersectionality shapes reproductive health experiences in ways that single-identity frameworks cannot capture. Five focus group participants emphasized how reproductive experiences emerge from multiple intersecting identities - race, class, disability status, sexuality, and gender identity (FG1-A, FG1-M, FG1-C, FG2-R, FG2-J). Medical racism and historical abuses have created justified distrust of Western medical institutions among some communities, yet health education rarely acknowledges this history. Effective reproductive health education must recognize how these identities interact to create unique barriers and experiences.

\subsection{Question-Based Approach: From Direct Correction to Self-Discovery}

All eight focus group participants strongly preferred question-based myth correction over direct statements, arguing that questions create psychological safety and enable self-discovery rather than triggering defensive reactions.

\subsubsection{Reducing Defensiveness Through Linguistic Framing}

Direct contradiction inhibits learning by triggering defensiveness, while carefully framed questions reposition learners as active participants in their own understanding (FG1-S, FG1-M). The linguistic architecture of questions proves as important as their content.

Effective questions employ several strategic framings. Opening with empathy and acknowledgment - phrasing such as "Many people believe X. Let's explore why that might not be accurate" - facilitates engagement. Collective language ("we") rather than accusatory language ("you") reduces feelings of individual targeting. Four focus group participants noted feeling less defensive when questions validated that misconceptions are widely shared rather than individual failings (FG1-M, FG1-A, FG1-C, FG2-J). This principle was echoed in user interviews, where six interview participants specifically commented on the neutral, non-judgmental tone of questions (P1, P2, P5, P7, P12, P13), with coded data revealing that tone and framing perception emerged as a concern across six interview excerpts.

Curiosity-driven questions using "why" and "how" formats prove more effective than binary evaluations, inviting exploration rather than judgment. Historical contextualization further reduces defensiveness by positioning myths as products of their historical context: "In the past, people believed X because of limited medical knowledge. What do we know now?" This framing treats myths as historically situated rather than indicators of ongoing ignorance.

Validation preceding correction emerged as particularly effective across all eight focus group participants. The recommended approach initiates exchanges by acknowledging concerns before introducing alternatives:

\begin{quote}
Your concern about [myth] makes sense given what's often shared on social media. Recent research actually shows... (FG2-S, FG2-R, FG2-J, FG2-L)
\end{quote}

This acknowledges that individuals form beliefs based on available information, even when that information proves flawed. Exploring why myths persist can prove as educational as learning accurate information, with questions like "Why do you think the belief that [myth] became so common?" inviting analytical thinking rather than defensive reactions.

\subsubsection{Designing Questions for Critical Reflection and Ethical Engagement}

Beyond basic myth correction, all eight focus group participants advocated for questions engaging values, scenarios, and open-ended inquiry that challenge stigma at structural levels rather than merely correcting individual misconceptions.

Scenario-based questions create productive distance from personal beliefs while maintaining relevance. Grounding questions in authentic situations - such as "Your friend thinks they might be pregnant but is afraid to take a test. What would you say?" - creates space to discuss sensitive topics without requiring personal disclosure. This psychological distance shifts focus from defending personal beliefs to navigating hypothetical situations. Interview data revealed ten coded excerpts related to engagement and discussion, with multiple participants noting that well-framed questions encouraged open discussion rather than defensiveness (P2, P9).

Values-based questions emerged as particularly powerful for revealing unstated assumptions. Three focus group participants praised this approach for challenging deeply held assumptions about gender roles by revealing contradictions in how individuals apply principles they already endorse (FG2-L, FG2-R, FG2-J). Examples included connecting reproductive health to broader principles: "If you believe in bodily autonomy, how does that apply to reproductive decisions?" and "Why do we treat fertility as a woman's responsibility when conception requires both partners?" 

Questions engaging emotion and empathy - addressing feelings and experiences rather than exclusively knowledge - promote sustained learning by engaging empathy alongside intellect. Open-ended questions without definitive correct answers encourage extended reflection beyond simple fact-verification.

However, four focus group participants cautioned against excessive ambiguity, emphasizing the need to balance complexity with clarity (FG1-M, FG1-A, FG1-C, FG2-R). The recommended approach involves scaffolding learning: beginning with accessible foundational questions requiring clear factual answers, then gradually increasing complexity through open-ended inquiry. As participants consistently emphasized, simplicity need not be condescending - questions can be accessible while treating individuals as capable of complex thought.

\subsection{Inclusive and Culturally Responsive Design for Reproductive Health Education}

Effective reproductive health education requires both scientific accuracy and cultural responsiveness. All eight focus group participants articulated specific strategies for inclusive design that centers marginalized voices while navigating the tension between cultural beliefs and evidence-based information.

\subsubsection{Language, Representation, and Centering Marginalized Identities}

Language choices signal belonging from the outset. Using "people who menstruate" rather than "women who menstruate" acknowledges that not all women menstruate and not all individuals who menstruate are women (FG2-L). Varying pronouns across questions (she/her, he/him, they/them) normalizes gender diversity rather than defaulting to binary assumptions.

Representation must extend beyond pronouns into names, scenarios, and explicit acknowledgment of structural barriers. Incorporating names from diverse cultural backgrounds - "Aisha," "Xiomara," "Priya," or "Kenji" - challenges defaults to culturally homogeneous examples (FG2-L). Economic barriers require explicit recognition, with one participant recommending reframing questions from "What birth control method would you choose?" to "If you had access to various birth control methods, what factors would influence your choice?" to acknowledge how access shapes reproductive decisions (FG2-S).

LGBTQ+ representation must be explicit rather than assumed. Three focus group participants argued that assuming users will "translate" scenarios to their own identities places unfair burden on marginalized groups, advocating instead for explicit inclusion of LGBTQ+ individuals in scenarios (FG2-J, FG2-L, FG2-R). Intentional inclusion counteracts the cumulative exclusions experienced through healthcare forms, clinical language, and educational materials.

Disability remains largely absent from reproductive health education despite disabled individuals having sexual and reproductive lives. Two focus group participants suggested scenarios featuring pregnant wheelchair users and individuals with sensory processing differences navigating pelvic exams, noting that disability is often completely erased from conversations about sexuality and reproduction (FG2-L, FG2-J). Digital accessibility - including alternative text, captions, and color contrast - must accompany representational inclusion.

Male inclusion emerged as necessary to challenge the framing of reproductive health as exclusively women's domain. Three focus group participants noted how men's systematic exclusion from conversations about contraception, fertility, and pregnancy support reinforces gendered assumptions, advocating for early education about men's role in reproductive health (FG2-S, FG2-J, FG2-R). Broadening reproductive health education beyond pregnancy to encompass menstruation, menopause, pleasure, and consent across life stages further challenges narrow framing.

\subsubsection{Balancing Cultural Sensitivity and Evidence-Based Information}

Cultural beliefs and scientific accuracy can coexist, though this balance requires strategies respecting both cultural meaning and medical evidence.

Many reproductive health myths hold cultural or religious significance, creating situations where correction can feel like heritage dismissal (FG2-L). However, three focus group participants clarified an important distinction: something can be scientifically incorrect while still holding personal or communal meaning (FG1-A, FG2-R, FG2-L). The hair-washing example discussed earlier illustrates this precisely - recognizing a belief as scientifically unfounded while continuing the practice culturally demonstrates how personal meaning coexists with scientific understanding.

Linguistic framing determines whether individuals feel validated or attacked. Starting with "This is a commonly held view..." rather than direct contradiction helps users feel validated (FG1-M). Using collective "we" language rather than accusatory "you" language creates shared inquiry, avoiding dynamics positioning the learner as wrong and the educator as superior (FG2-R).

Acknowledging multiple epistemologies prevents positioning Western science as the sole valid knowledge system. Three focus group participants emphasized that scientific knowledge represents one epistemology while cultural knowledge, traditional practices, and lived experience constitute equally valid forms of understanding (FG2-S, FG2-R, FG2-J). Framing information as "here's what medical research shows" rather than "here's the truth and everything else is false" provides evidence-based information without dismissing other knowledge systems.

Avoiding targeting specific cultural or religious groups emerged as critical across all eight focus group participants. Examples should be generalized or drawn from multiple contexts rather than singling out particular communities (FG2-L). The application must clearly communicate that myth correction aims to provide health information rather than judge cultural practices or individual beliefs (FG2-J).

\subsection{AI Limitations and the Essential Role of Human Oversight}

All eight focus group participants demonstrated sophisticated understanding of AI's potential alongside significant concerns about its limitations for reproductive health education, revealing why human oversight constitutes an essential requirement rather than optional enhancement.

AI's tendency toward hallucination poses particular dangers for health information. While acknowledging AI's pattern recognition capabilities, four focus group participants emphasized critical limitations: AI can generate plausible-sounding but inaccurate information, particularly dangerous for health topics where users might trust authoritative-sounding content (FG2-R, FG2-J, FG2-S, FG2-L).

Training data limitations risk perpetuating existing health inequities. Three focus group participants expressed concern that if reproductive health research has historically centered certain populations while excluding others, AI trained on this data would perpetuate rather than address these gaps (FG2-S, FG2-R, FG2-J). This creates particular weaknesses in areas like LGBTQ+ reproductive health where limited training data exists (FG2-J).

Culturally contested information presents unique challenges for AI systems. Four focus group participants identified the difficulty of navigating situations where medical consensus exists but cultural or religious beliefs differ, warning that AI might incorrectly label culturally significant beliefs as misinformation (FG2-L, FG2-R, FG2-J, FG2-S). The challenge extends beyond identifying misinformation to responding appropriately - AI might correct myths in culturally insensitive or dismissive ways (FG2-J). Additionally, "it depends" sometimes represents the most accurate answer, yet AI might default to more definitive statements that oversimplify nuanced situations (FG2-R).

Human oversight emerged as essential across all focus group participants. Recommendations included pilot testing content with diverse communities before widespread release and - more fundamentally - involving community members in content development rather than merely testing final products (FG2-R, FG2-S). One participant emphasized the need for ongoing evolution:

\begin{quote}
What feels inclusive today might feel inadequate in five years as understanding evolves. (FG2-L)
\end{quote}

This recognition positions inclusive design as an ongoing process requiring sustained human engagement rather than a one-time technical solution.

\subsection{User Evaluation: The Implementation Gap}

User testing of the prototype application revealed substantial disconnect between the sophisticated design principles participants articulated in focus groups and the actual output generated by the AI system, demonstrating current limitations in deploying large language models for stigma reduction in reproductive health education.

Interview data revealed nineteen coded excerpts related to question quality and depth, with participants consistently noting that AI-generated questions felt superficial, merely rephrasing article content rather than encouraging critical thinking. One participant captured this limitation:

\begin{quote}
It wasn't actually adding more thought or like follow up discussion. It really was like rephrasing a sentence in a different way to make it a question... it just like rephrasing sentences like literally exact sentences in the article... it feels like it was more of like an attention check than actually like spurring conversation or deep thought.
\end{quote}

Content relevance issues compounded quality concerns, with questions often addressing trivial details rather than meaningful health information. Thirteen coded excerpts addressed content relevance and accuracy concerns. Multiple participants noted questions about article metadata - authors, website extensions - rather than substantive content (P1, P3). Accuracy issues emerged as well, with participants reporting questions about statistics that contradicted article information or offered answer choices that didn't align with stated facts.

The AI defaulted to factual recall rather than the critical reflection participants deemed essential. While some factual questions provide necessary foundational knowledge, interview participants consistently noted an imbalance heavily favoring statistics and facts over individual perspectives, open discussion, and intellectual engagement (P3, P7, P8).

However, cultural sensitivity received positive evaluations across interview participants. Eight coded excerpts confirmed that testers found the content culturally appropriate, affirming that questions were neutral and language did not evoke feelings of insensitivity. All identified interview participants confirmed finding nothing insensitive or inappropriate (P1, P2, P3, P4, P5, P6, P7, P8, P9). This suggests that current large language models can avoid overtly offensive language, meeting this baseline requirement.

Multimedia enhancement emerged as an opportunity, with five coded excerpts suggesting that infographics, graphics, and images would significantly improve engagement and comprehension (P4, P5).

Technical functionality presented mixed results. Ten coded excerpts addressed user interface and experience, with multiple participants praising the application's simplicity and ease of navigation (P2, P3, P4, P5, P6). However, nine coded excerpts documented technical issues including failed URL submissions, loading errors, answer resets when scrolling, and parsing problems (P1, P2, P9).

The implementation gap proves clear: while focus group participants articulated sophisticated approaches emphasizing values-based reflection, cultural nuance, and open-ended inquiry, the AI defaulted to surface-level factual questions that merely rephrased article content. The AI succeeded at avoiding overtly offensive language but failed at promoting the critical reflection essential for stigma reduction. This gap reinforces the necessity of human oversight in reproductive health education technology - technical competence at avoiding offense does not equate to pedagogical effectiveness at transforming deeply held beliefs and challenging systemic stigma.

\subsection{Quantitative Results: Post-Use Survey}

After interacting with the AI-generated questions in the application, participants completed a survey with 12 questions assessing their perceptions of the system. Each item used a 5-point Likert scale (1 = Strongly Disagree, 5 = Strongly Agree). We report both response counts and average scores for each survey item.

\begin{figure}[H]
    \centering
    \includegraphics[width=\textwidth]{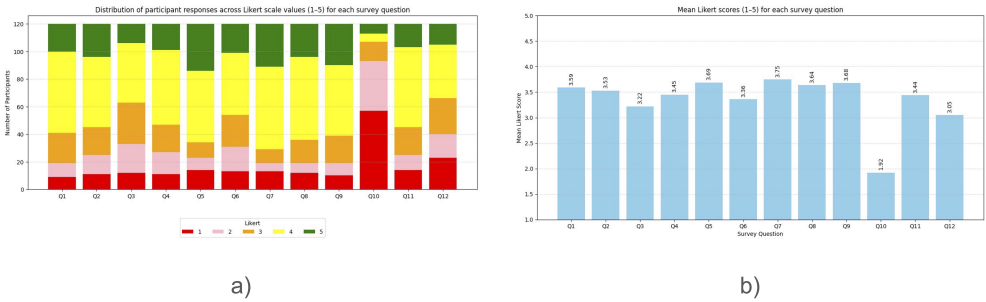}
    \caption{Survey results: (a) Frequency of responses for each survey question, (b) Mean Likert scores (1–5).}
    \label{fig:survey_graphs}
\end{figure}

Figure~\ref{fig:survey_graphs} summarizes the survey results for all 12 questions. Panel (a) shows the distribution of participant responses across Likert scale values (1–5) for each survey question. Panel (b) presents the mean Likert score for each survey question (1 = Strongly disagree, 5 = Strongly agree). All participants (N = 120) responded to all questions.

Overall, the survey results indicate generally positive participant perceptions of the AI-generated questions. Across most items (Q1–Q9, Q11–Q12), responses were skewed toward Agree and Strongly agree, with mean Likert scores clustering above the neutral midpoint, suggesting that participants found the questions engaging, relevant, and effective in promoting critical thinking about reproductive wellbeing. In particular, high agreement was observed for items related to relevance, societal importance, and engagement. In contrast, Q10, which assessed whether any questions were perceived as offensive or inappropriate, showed a notably lower mean score and a higher concentration of disagreement responses, indicating that most participants did not find the language or framing objectionable. Together, these results suggest that the AI-generated questions were largely well-received while maintaining sensitivity to potentially delicate topics.

%% file: 5.discussion.tex
\section{Discussion}

This study examined tensions that emerge when AI-generated, question-based content is deployed to engage with reproductive wellbeing topics amid stigma, cultural sensitivity, and inclusion challenges. By combining qualitative insights with post-use survey results, our findings reveal both the promise and current limitations of LLMs for stigma-aware reproductive health education.

\subsection{Design Implications}

Our findings generate several design implications for human–AI interaction (HAI) in stigmatized health domains and for culturally sensitive AI design more broadly. We found that reproductive health stigma is not simply a consequence of information scarcity, but a deeply social and cultural phenomenon shaped by moral norms, institutional practices, and historically situated narratives. Although participants generally perceived the AI-generated questions as relevant and non-offensive, they emphasized that stigma persisted when interactions failed to engage values, lived experience, and social meaning. This aligns with prior work showing that stigma endures even when factual knowledge is available, such as in menstrual health education and contraceptive awareness campaigns~\cite{tuli2019girl, eshak2020myths}. For designers, this suggests that access to information alone is insufficient without attending to how knowledge is framed and interpreted within cultural contexts.

Participants consistently articulated a preference for AI interactions that support reflection rather than information recall. They valued question framing that acknowledged existing beliefs, employed empathetic and inclusive language, and situated myths within social or historical contexts. This points to a design opportunity for AI systems that foreground reflective, values-based questioning instead of neutral fact delivery. Prior HCI and public health research has shown that question framing influences engagement and learning outcomes~\cite{wang2024exploring, cook2014reducing}. Our findings extend this work by demonstrating how such framing can reduce defensiveness and encourage dialogue in stigmatized reproductive health settings. Designers should therefore treat how AI asks questions as a central design concern, rather than a secondary linguistic choice.

Language inclusivity and representation emerged as critical design considerations. Participants emphasized the need for AI-generated content to account for diverse gender identities, family structures, and relational roles in reproductive decision-making. Concerns about the marginalization of trans and non-binary individuals, as well as the exclusion of men from reproductive health discourse, echo broader critiques of reproductive health systems that rely on narrow normative assumptions~\cite{kukura2022reconceiving, almeling2014more}. These findings suggest that stigma-aware AI design requires deliberate attention to who is represented, who is addressed, and whose perspectives are implicitly centered or excluded in AI-generated interactions.

Finally, the implementation gap observed in our prototype highlights the limits of current LLM capabilities and the necessity of human-guided scaffolding. While the LLM reliably avoided overtly offensive language, it often defaulted to superficial rephrasing rather than facilitating critical inquiry. This underscores the importance of participatory and human-in-the-loop design approaches, where communities, educators, and domain experts actively shape prompt structures, content boundaries, and evaluative criteria. Consistent with prior work showing that human-facilitated health education often outperforms AI-only approaches in sensitive contexts~\cite{kim2025human, kelly2024chatbot}, we argue that LLMs should be positioned as supporting tools rather than autonomous educators in reproductive health. Such an orientation aligns with DIS commitments to responsible, culturally grounded, and socially accountable HAI design.

\subsection{OpenBloom as Feminist Interaction Design}

Our analysis revealed that participants’ interactions with OpenBloom simultaneously opened spaces for curiosity, reflection, and engagement with reproductive wellbeing, while also surfacing discomfort around sensitivity and care. Some participants described the AI-generated questions as inviting them to think about reproductive health in ways they had not previously encountered, creating moments of openness and dialogue. At the same time, we observed persistent tensions stemming from superficial question framing, limited cultural contextualization, and the absence of empathetic engagement, which often constrained deeper reflection and reinforced existing stigmas rather than challenging them. These moments of possibility and breakdown echo broader sociocultural struggles surrounding reproductive health education, where efforts to increase access to information coexist with deeply rooted norms of silence and exclusion.'

To make sense of these dynamics and articulate their broader design implications, we turn to Bardzell’s Feminist HCI framework~\cite{bardzell2010feminist}. The framework’s six interrelated qualities—pluralism, participation, advocacy, ecology, embodiment, and self-disclosure—offer a productive lens for understanding not only what OpenBloom enabled, but also where current LLM-mediated interactions fall short in supporting stigma-sensitive and culturally grounded reproductive health engagement. Through this lens, OpenBloom functions as a design probe that surfaces the interactional limits of existing AI systems and points toward future feminist-oriented design directions for reproductive health technologies.

\subsubsection*{Pluralism and the Implementation Gap.} Bardzell defines pluralism as rsistance to singular, totalizing, or universal viewpoints in design. OpenBloom's question-based approach attempted pluralism by transforming articles into prompts for exploration rather than delivering authoritative corrections. However, our findings reveal a critical implementation gap: while participants articulated sophisticated preferences for questions that acknowledge diverse family structures, gender identities, and cultural contexts (Section 4.2), the LLM defaulted to relatively generic outputs. The AI struggled to generate questions reflecting the "constellation of marginal perspectives" that Bardzell argues characterizes pluralist design. For instance, participants noted how questions about contraception assumed autonomous decision-making without acknowledging family-centered dynamics common in many cultural contexts. This gap demonstrates that achieving pluralism requires more than avoiding offensive language—it demands actively centering diverse experiences and questioning whose knowledge counts as legitimate. Future iterations might address this through participatory prompt development where communities contribute to defining what pluralist reproductive health questions look like in their contexts.

\subsubsection*{Advocacy Without Imposition.} Bardzell frames advocacy as engaging with political emancipation while questioning designers' authority to define "improved society." OpenBloom embodied this tension: we sought to challenge reproductive health stigma (advocacy) while using participant feedback to shape what appropriate challenges look like (questioning our authority). Our findings suggest participants valued this approach—they consistently preferred questions that validated existing knowledge before introducing alternatives, and emphasized the importance of human oversight to prevent AI from imposing values (Section 4.1). However, the implementation gap again limited advocacy's realization. Questions that might challenge stigma through careful values-clarification or scenario-based reasoning—approaches participants described as desirable—were largely absent from AI outputs. This reveals a design challenge for feminist AI: how do we build systems that advocate for progressive change (challenging reproductive health myths that harm people) while remaining accountable to the communities they serve? Our findings suggest this requires hybrid approaches where humans explicitly scaffold AI outputs toward advocacy goals that emerge from community dialogue rather than designer assumptions.

\subsubsection*{Embodiment and Stigmatized Health Topics.} Bardzell emphasizes embodiment as foregrounding gender, sexuality, pleasure, emotion, and desire in interaction design—moving beyond disembodied mental models toward recognition that interactions engage whole bodies with drives and experiences. Reproductive health is fundamentally embodied, yet paradoxically this embodiment is often what generates stigma. Our findings show participants were acutely aware of how questions might make them feel vulnerable, defensive, or ashamed—affective responses tied to embodied reproductive experiences often deemed inappropriate for public discussion (Section 4.2). Participants' preferences for empathetic language, collective framing ("many people experience"), and historical contextualization all represent strategies for addressing embodiment without amplifying stigma. This suggests that effective feminist AI for reproductive health must explicitly account for how questions land on bodies marked by gender, marked by reproductive capacity or its absence, marked by medical histories that may be traumatic or surveilled. The question-based format showed potential here—questions can invite reflection about embodied experiences without demanding disclosure—but realizing this potential requires more sophisticated prompt engineering that keeps embodiment and affect central throughout content generation.

\subsubsection*{Self-Disclosure and AI Transparency.} Bardzell's quality of self-disclosure refers to making visible how software constitutes users as subjects—revealing the assumptions built into systems and creating space for users to redefine themselves against those assumptions. OpenBloom attempted this through transparency about using AI and soliciting feedback, but our findings suggest opportunities for deeper self-disclosure. Participants expressed strong awareness of AI limitations, identifying risks of hallucination, training data bias, and cultural oversimplification (Section 5.2). Rather than undermining trust, this awareness prompted valuable reflections about what kinds of health questions require human expertise versus where AI might assist. Future designs might embrace this by making AI's assumptions about reproductive health more visible within the interaction itself—for example, explicitly noting when questions draw from Western medical frameworks, or flagging when cultural context matters that the AI cannot access. This self-disclosure could transform limitations into learning opportunities, helping users develop critical literacy about AI-mediated health information while acknowledging the system's bounded perspective rather than claiming false universality.

\subsubsection*{Participation and Ecology: Future Directions.} While OpenBloom did not fully realize participatory or ecological qualities, our findings point toward how future iterations might. Participation—valuing collaborative processes where users and designers create together—emerged as essential in participants' emphasis on human oversight and community-specific adaptation (Section 4.3). Participants effectively told us that reproductive health education cannot be designed \textit{for} communities without designing \textit{with} them, given how profoundly cultural contexts shape what questions feel appropriate or offensive. The ecological quality—attending to artifacts' effects across broadest contexts—connects to participants' concerns about privacy, surveillance, and how reproductive health technologies operate amid restrictive legal environments (Section 5.2). Both qualities suggest that scaling question-based AI for reproductive health should not mean deploying universal templates, but rather developing participatory methods where communities shape local implementations while attending to the broader sociotechnical ecologies these systems enter.

Viewed through a Feminist HCI lens, our findings suggest that the technical ability to generate grammatically correct or non-offensive questions is necessary but far from sufficient. Feminist HCI foregrounds commitments to pluralism, advocacy, embodiment, self-disclosure, participation, and ecology—qualities that extend beyond surface-level language safety to the social relations and power dynamics embedded in interaction. From this perspective, OpenBloom functions not as a completed intervention but as a design probe that surfaces where current LLM capabilities misalign with these feminist commitments. By empirically revealing how AI-generated questions fall short of participants’ expectations for care, inclusivity, and reflexivity, our work demonstrates why reproductive health AI demands explicitly feminist design approaches and clarifies the limits of purely technical solutions in this area.

\subsection{Broader Implications}

Our findings reinforce that inclusive reproductive health education requires more than neutral AI systems. Participants emphasized that healthcare and educational technologies frequently exclude marginalized communities through assumptions in language, examples, and system design. Because of this, inclusive AI design must be intentional, by explicitly representing gender diversity, varied cultural contexts, and accessibility needs rather than assuming users will adapt content to themselves. These implications are important in the current reproductive rights landscape. Recent HCI research has shown how reproductive health organizations increasingly rely on digital tools to navigate access, safety, and information sharing in the aftermath of Roe v. Wade ~\cite{petterson2025protected}. At the same time, reproductive privacy has become a concern involving platforms, policies, algorithms, and users themselves ~\cite{karizat2025laboring}. Our findings suggest that LLM-driven educational systems must be designed with these broader legal, political, and privacy contexts in mind, especially when deployed around stigmatized and potentially surveilled forms of care.

These insights relate to previous research on culturally responsive AI and health communication ~\cite{davies2024culturalsensitivity} ~\cite{who2024ai}, while extending this work by showing how exclusion happens not only through misinformation, but also through omission and default assumptions. Our study shows a tension between scalable, standardized AI templates and the need for culturally nuanced, intersectional content. Navigating this tension requires positioning AI as a collaborator, with humans actively shaping content, scenarios, and framing. Situating our findings within a reproductive justice framework further highlights how algorithmic systems can unintentionally reinforce inequities if not carefully designed ~\cite{fledderjohann2024algorithmic}. Prior work demonstrates that reproductive health experiences are deeply shaped by cultural, religious, and social norms across global contexts, including among Arab Muslim communities ~\cite{alnaimi2024cultural}, rural and married populations in South Asia ~\cite{ghule2018barriers, casterline2001obstacles}, and individuals managing covert contraceptive use due to family or partner opposition ~\cite{akoth2021covert}. Male partner resistance to contraception remains a significant barrier, with women and providers navigating complex dynamics where anticipated or experienced resistance affects care-seeking behaviors and service delivery ~\cite{britton2021male}. Research on male involvement in reproductive care reveals that education, urban residency, and media exposure are associated with greater engagement, showing the importance of addressing systemic factors that shape partner participation ~\cite{bishwajit2017male}. Knowledge gaps among adolescents about reproductive health persist across diverse settings ~\cite{hossain2017knowledge}, highlighting the need for age-appropriate and culturally sensitive educational interventions. Additionally, the critical role of menstrual health in broader reproductive wellbeing is increasingly recognized ~\cite{critchley2020menstruation}. These structural barriers underscore that AI systems must be adaptable to diverse realities rather than assuming autonomy or openness in reproductive decision making.

Furthermore, participants showed strong awareness of AI limitations, identifying hallucination risks, bias in training data, and oversimplification of cultural context. Consistent with prior work ~\cite{hanai2024generative} ~\cite{mills2023chatbots}, our findings reinforce that human oversight is not optional but rather essential in reproductive health applications. AI can assist in content generation and accessibility, but human judgment is required to ensure cultural sensitivity, ethical framing, and effectiveness. Evidence from prior work support this conclusion. Studies evaluating generative AI chatbots for youth mental health reveal similar limitations around contextual understanding and appropriateness in sensitive domains ~\cite{sobowale2025evaluating}. Additionally, digital health research highlights that effective behavior change depends on techniques such as reflection, personalization, and social framing rather than just information delivery ~\cite{mair2023behavior}. Broader health equity research emphasizes that respectful, person-centered care is fundamental to positive maternal and reproductive health outcomes, particularly in low and middle-income contexts where systemic barriers compound individual challenges ~\cite{kawish2023respectful}.
Together, we contend that the broader impact of AI in reproductive health to depend not on scale alone, but also on whether systems are designed to support respect and trust across diverse populations.

\subsection{Limitations and Future Work}

Our work has several limitations. First, while our multi-method approach provides insight, our sample size and recruitment methods limit generalizability. Second, our prototype relied on relatively simple prompting strategies, which constrained the depth of AI-generated questions. Future work should explore systematic prompt creation and human involved workflows that prioritize reflection, empathy, and stigma awareness.

Additionally, future research can build on insights from prior reproductive health technology studies. For example, menstrual health research shows the need to consider underrepresented populations and social contexts ~\cite{hennegan2019menstruation, epstein2017tracking, bardzell2018menopause}. Participatory and inclusive design approaches have been recommended for menopause ~\cite{backonja2021menopause, tutia2018hci} and menstrual tracking technologies ~\cite{lin2023functional, tuli2021rethinking, mirzaliyeva2024ai, tuli2019immobilities}, while reproductive health interventions in workplace and maternal contexts emphasize community-led interventions and AI support tools ~\cite{ng2019entrepreneur, kumar2015projecting, joshi2013family, tumpa2017smartcare}. Future research can examine how different populations experience AI-generated reproductive health content. Longitudinal studies may further assess whether reflective, question-based AI systems can meaningfully shift attitudes or reduce stigma over time. Finally, participatory design approaches that engage community members in contributing to AI prompts and content are a promising direction for building more inclusive and trustworthy solutions.

%% file: 6.conclusion.tex
\section{Conclusion}
This paper examined how LLM-generated, question-based educational content engages with reproductive wellbeing in stigmatized and culturally sensitive contexts. Using OpenBloom as a design probe, we conducted interviews, surveys, and focus group discussions to investigate how participants interpret the sensitivity, educational value, and cultural appropriateness of AI-generated questions. Our findings reveal a gap between participants’ expectations for stigma-aware, reflective educational engagement and the interactional patterns produced by current LLMs. While AI-generated content generally avoided overtly offensive language, it often relied on superficial reframing and factual recall rather than the empathetic, values-oriented questioning participants described as necessary for engaging stigma. This work contributes design knowledge to HAI by articulating key tensions in designing stigma-sensitive AI, including question framing, representation, and the role of human oversight. We argue that participatory, culturally grounded design processes are essential when deploying LLMs in reproductive health education, where AI can support human expertise and community knowledge.

%% file: 7.appendix.tex
\appendix

\section{Survey Questions}

The following Likert-scale questions were presented to participants
(1 = Strongly disagree, 5 = Strongly agree).

\begin{enumerate}
    \item These questions challenged me to think critically about reproductive wellbeing.
    \item These questions were interesting and engaging.
    \item These questions presented reproductive wellbeing topics in a unique and creative way.
    \item I resonated with the themes of the questions.
    \item These questions provided me with new knowledge or perspectives on reproductive wellbeing.
    \item These questions increased my awareness and motivation to challenge stigma around reproductive wellbeing.
    \item These questions accurately reflected the key themes of the article.
    \item These questions are relevant to our social norms.
    \item These questions addressed issues that are important in today’s society.
    \item Some questions contained language or framing that I found potentially offensive or inappropriate.
    \item These questions make me feel comfortable discussing reproductive wellbeing topics.
    \item I found these questions to be sensitive to different cultural backgrounds.
\end{enumerate}

\section{Interview Guide}

\subsection*{Creativity \& Cultural Sensitivity}
\begin{enumerate}
    \item Do the questions feel engaging and thought-provoking?
    \item Do they challenge stigma appropriately within cultural contexts?
    \item Do the questions reflect diverse perspectives and lived experiences?
    \item Are the questions framed in a way that encourages open discussion rather than discomfort or defensiveness?
    \item Do the questions adapt well to different cultural and social backgrounds?
\end{enumerate}

\subsection*{Approach to Stigma}
\begin{enumerate}
    \item How well do the questions address and help break down stigma?
    \item Are there any aspects that feel insensitive or inappropriate?
    \item Do the questions encourage critical thinking about social norms and taboos?
    \item Do the questions provide enough factual information to correct misconceptions?
    \item How comfortable did you feel discussing your responses to the questions?
\end{enumerate}

\subsection*{Usability \& Experience}
\begin{enumerate}
    \item Was the app user-friendly?
    \item Was the language accessible and inclusive?
    \item Any technical or content-related issues?
    \item Did you find the app’s interface intuitive and easy to navigate?
    \item Was the pacing of the questions appropriate for engagement and reflection?
    \item Did the app provide adequate explanations or resources to support understanding?
\end{enumerate}

\subsection*{Areas for Improvement}
\begin{enumerate}
    \item Any suggestions for refining the AI-generated questions?
    \item Additional features that could enhance the learning experience?
    \item Would multimedia elements (such as videos or infographics) improve the learning experience?
    \item Would you like more personalization options, such as tailoring questions to specific age groups or cultural backgrounds?
    \item Did you feel there were any gaps in the topics covered that should be included?
\end{enumerate}

\section{Focus Group Protocols}

\subsection*{Thought Provoking}
\textbf{Goal:} Explore what makes a question engaging or discussion-worthy rather than just fact recall.
\begin{enumerate}
    \item What stigmas do you believe need thought-provoking discussions?
    \item When you think about a "thought-provoking question", what comes to mind? Can you share an example?
    \item How open-ended should questions be? Should they have a "right" answer, or invite multiple perspectives?
    \item Some users found questions too simple, while others liked the mix. What’s the right balance between challenge and accessibility?
    \item In your experience, what makes you stop and really think about a question instead of just recalling facts?
    \item If we wanted the app to spark discussion between peers, what types of questions would be most effective?
\end{enumerate}

\subsection*{Myth Busting}
\textbf{Goal:} Understand if AI can challenge misconceptions without sounding judgmental or offensive.
\begin{enumerate}
    \item What are some myths that need to be busted?
    \item In your view, what role should questions play in addressing common myths or misinformation around reproductive wellbeing?
    \item How can we frame myth-busting questions so they feel constructive and not confrontational?
    \item From your technical/engineering perspective: Do you think AI can effectively detect and respond to misinformation? What are potential risks?
    \item What kinds of wording or framing would make you more receptive to a myth-busting question?
    \item Do you think myth-busting could unintentionally cause offense, and if so, how could we prevent that?
\end{enumerate}

\subsection*{Inclusivity}
\textbf{Goal:} Go beyond "not offensive" to identify actionable ideas for inclusive design.
\begin{enumerate}
    \item How can we ensure that questions reflect diverse gender identities, experiences, and cultural perspectives in reproductive wellbeing?
    \item What kinds of examples, pronouns, or scenarios would make questions feel more inclusive?
    \item Can you think of times when health or education tools felt exclusive or overlooked certain groups? What should we avoid repeating?
\end{enumerate}